\documentclass[%
 aip,
 amsmath,amssymb,
 reprint,
 twocolumn,
 floatfix
]{revtex4-1}

\usepackage{amsmath} 
\DeclareMathOperator{\erf}{erf}

\usepackage[english]{babel}
\usepackage{color}
\definecolor{Blue}{rgb}{0.3,0.3,0.9}  
\usepackage{graphicx}
\usepackage{dcolumn}
\usepackage{bm}
\usepackage{hyperref}
\hypersetup{
 colorlinks=true,
    linkcolor=blue,
    citecolor=blue,      
    urlcolor=blue,
}
\usepackage{mdframed}
\usepackage{todonotes}
\graphicspath{{Figures/}}
\usepackage{comment}
\usepackage{xcolor}
\usepackage{tikz}
\usetikzlibrary{positioning}
\newcommand{\COMMENT}[1]{}
\begin{document}

\title{Lithium Droplet Transport in Tokamak Edge Plasmas
}

\author{A. Diaw }
\email{diawa@ornl.gov}
\author{J.D. Lore}
\author{S. Smolentsev }
 \affiliation{Oak Ridge National Laboratory, Oak Ridge, TN 37932 U.S.}
            
\date{\today}

\begin{abstract}

A lithium droplet transport and evaporation model has been developed within the Direct Simulation Monte Carlo code OpenEdge. This model integrates gravity, collisional ion drag, orbital-motion-limited charging, energy-balance evaporation, and an anisotropic rocket recoil force using a Strang-split integrator. Validation against analytical drag-gravity solutions and independent RK45 evaporation integration demonstrates relative errors below $10^{-5}$ for droplet radii of 1.5, 2.5, and 3.5~mm. Simulations of ensembles containing $10^5$ droplets, launched from inner and outer divertor surfaces in SOLPS-ITER plasma background for the CAT tokamak reactor concept, indicate that transport outcomes are determined by initial size, velocity, and launch location. Outer-divertor droplets predominantly redeposit locally, whereas inner-divertor droplets reach the low-field-side wall. Smaller droplets lose most of their mass to evaporation before reaching the core, while larger droplets retain their mass and redeposit on nearby tiles. Both one-way and iterative two-way coupling frameworks map the evaporated lithium onto the SOLPS-ITER mesh as volumetric sources, facilitating self-consistent evaluation of lithium droplet impacts on edge-plasma performance.

\end{abstract}

\maketitle

\section{\label{sec:intro} Introduction}

\footnotetext{This manuscript has been authored in part by UT-Battelle, LLC, under contract DE-AC05-00OR22725 with the US Department of Energy (DOE). The publisher acknowledges the US government license to provide public access under the DOE Public Access Plan (http://energy.gov/downloads/doe-public-access-plan).}

Plasma-facing components (PFCs) in reactor-relevant tokamaks are required to withstand sustained high heat and particle fluxes~\cite{StangebyPPCF2022,Roth2008PPCF}. Liquid-lithium limiter and divertor concepts provide self-healing surfaces, reduced radiative losses, and enhanced vapor shielding~\cite{Maingi2011}. Experimental campaigns on CDX-U, NSTX, and EAST have demonstrated improved heat-flux management and impurity control with flowing lithium~\cite{TANG2020100845}. However, these studies also identified a limiting mechanism: liquid-surface instability and splashing. Under transient and magnetohydrodynamic (MHD) forcing, ejected lithium droplets enter the edge plasma, fragment, evaporate, and alter local plasma conditions~\cite{TANG2020100845}. Quantitative assessment of this process is necessary for predictive divertor evaluation.

Conventional edge-plasma source models do not explicitly account for droplet transport. Droplets, as finite-mass entities, exhibit coupled trajectory and thermal evolution that determine whether lithium remains localized to the divertor, redeposits on adjacent tiles, or contributes to distributed plasma sources. Recent liquid-metal (LM) divertor SOLPS studies have addressed geometric and source sensitivity but have not incorporated droplet-trajectory physics~\cite{Islam2022NME,Lore2022TPS}. Further studies with boundary transport codes have considered the impact of plasma drifts \cite{Islam2025}, and vapor shielding \cite{Marenkov_2021, Marenkov_2022, Emdee_2024}, redeposition \cite{Marenkov_2021}, and hydrogen retention \cite{Marenkov_2025, Morbey_2024}. 

This study addresses this gap by developing a lithium-droplet model analogous to those used for dust transport in tokamaks. The model is implemented and verified in OpenEdge~\cite{OpenEdge}, and subsequently applied within a fixed SOLPS plasma background for the CAT fusion pilot plant concept \cite{Buttery_2021} to quantify size- and launch-dependent fate statistics.

An iterative two-way coupling framework is developed to exchange lithium sources and updated plasma profiles between OpenEdge and SOLPS-ITER, employing under-relaxation \cite{Kuttler08,Zhao2019} to ensure stability. 


The structure of this paper is as follows. Section~\ref{sec:DropletModel} presents the droplet model. Section~\ref{sec:Implementation/Tests} describes the numerical scheme and verification tests. Section~\ref{sec:Results} reports parametric studies, and the two-way coupling implementation, and discusses implications for liquid-metal divertor design.

\section{Droplet Model Formulation}
\label{sec:DropletModel}
\subsection{Droplet-particle parameters}
Each droplet is represented as a spherical particle characterized by radius $r_d$, uniform density $\rho_d$, and temperature $T_d$. The geometric and material properties are summarized below:
\begin{align}
V_d &= \frac{4}{3}\pi r_d^{3}, \\
m_d &= \rho_d V_d, \\
\sigma_d &= \pi r_d^{2}, \\
N_d &= \frac{m_d}{m_a} = \frac{\rho_d V_d}{m_a},
\end{align}
where $m_a$ is the atomic mass of lithium, $m_d$ is the droplet mass, and $\sigma_d$ is its effective cross-sectional area.

Initial conditions are based on parameters reported in the LM divertor ejection literature, with diameters ranging from approximately $20~\mu\mathrm{m}$ to $1~\mathrm{cm}$, velocities up to $20~\mathrm{m\,s^{-1}}$, and shallow launch angles~\cite{TANG2020100845}. Within this parameter space, mist, droplet, and ligament regimes are identified. The parametric analysis examines three representative initial radii: $1.5$, $2.5$, and $3.5~\mathrm{mm}$, which are applied consistently throughout the study.

\subsection{Lithium droplet dynamics in a plasma}
\label{sec:equationofMotion}

A spherical droplet with mass $m_d$, charge $q_d$, and radius $r_d$ is considered. The analysis adopts the dust-particle model described in \cite{DUSTT}, which incorporates mass loss due to evaporation. The equations of motion are as follows:
\begin{eqnarray}
\label{eq:xdot}
\frac{d\mathbf{x}}{dt} &=& \mathbf{v}_d, \\[4pt]
\label{eq:vdot}
\frac{d}{dt}\!\left(m_d \mathbf{v}\right)
&=&  \mathbf{F}_{\rm d}.
\end{eqnarray}
The total force is
\begin{equation}
 \mathbf{F}_{\rm d} =
 \mathbf{F}_{\rm drag, i} + \mathbf{F}_E + \mathbf{F}_g,
\end{equation}
with
\begin{eqnarray}
\mathbf{F}_{\rm drag,i}
&=&
\xi_i\,\xi_{\rm fric,i}\,
m_i n_i v_{T_i}\,
\bigl(\mathbf{u}_i - \mathbf{v}_d\bigr)\,
\sigma_d,
\\[4pt]
\mathbf{F}_{\rm E}
&=&
-\,e Z_d\,\xi_E\,\mathbf{E}_{\rm plasma},
\\[4pt]
\mathbf{F}_{\rm g}
&=&
m_d\,\mathbf{g}.
\end{eqnarray}
In this context, $m_i$, $n_i$, and $\mathbf{u}_i$ denote the mass, density, and flow-velocity vector of the plasma ions, respectively. The ion thermal velocity is defined as $v_{T_i}=(2T_i/m_i)^{1/2}$. The term $\mathbf{E}_{\rm plasma}$ represents the electric field in the plasma, and $\mathbf{g}$ denotes gravitational acceleration. The coefficients $\xi_i$ and $\xi_E$ serve as scale factors, which may be used to account for deviations from spherical droplet geometry.

Analytical expressions for the friction force $\mathbf{F}_{\rm drag,i}$ between droplet and plasma ion particles are given in Refs.~\cite{Hutchinson05,2004PhRvE..70e6405K}. The force has two components: $\mathbf{F}_{\rm drag,i} = \mathbf{F}_{\rm coll} + \mathbf{F}_{\rm orb}$, 
where $\mathbf{F}_{\rm coll}$ is due to direct ion collection by the droplet, and $\mathbf{F}_{\rm orb}$ is due to Coulomb scattering.

For a negatively charged sphere, the first component reads as 
\begin{eqnarray*}
\mathbf{F}_{\rm coll}
&\!=\!&
\frac{\mathbf{F}_{\rm Epstein}}{2 u^{3}\sqrt{\pi}}
\Bigg\{
u\left(2u^{2}+1+\frac{2\chi}{\delta_{\rm ite}}\right)e^{-u^{2}}
\nonumber\\[4pt]
&&\qquad
+\,\frac{\sqrt{\pi}}{2}
\left[
4u^{4}-1-2\left(1-2u^{2}\right)\frac{\chi}{\delta_{\rm ite}}
\right]\erf(u)\Bigg\}.
\end{eqnarray*}
Here $u = |\mathbf{u}_i-\mathbf{v}_d|/v_{T_i}$ is the normalized ion–droplet relative speed, $\chi$ is the Coulomb parameter,  $\delta_{\rm ite}$
denotes the ion–thermal energy ratio, and the Epstein force is given by:
\begin{equation}
\mathbf{F}_{\rm Epstein}
= m_i n_i v_{T_i}(\mathbf{u}_i-\mathbf{v}_d)\sigma_d .
\end{equation}

The orbital-scattering component has the form
\begin{equation}
\mathbf{F}_{\rm orb}
=
2\,\mathbf{F}_{\rm Epstein}
\left(\frac{\chi}{\delta_{\rm ite}}\right)^{2}
\ln\Lambda\,
\frac{\mathcal{Y}(u)}{u},
\end{equation}
where the Chandrasekhar function is
\begin{equation}
\mathcal{Y}(u)
=
\erf(u) - \frac{2u}{\sqrt{\pi}}e^{-u^{2}},
\end{equation}
and $\ln\Lambda$ is the Coulomb logarithm. 

For lithium droplets ranging from micron to millimeter scale in divertor plasmas, the Lorentz force is negligible compared to gravity and collisional drag. Although the Lorentz force is included for completeness, gravity and drag are the primary forces governing the droplet dynamics.

\subsection{Droplet charging model}
\label{sec:droplet_charging}

A droplet immersed in the edge plasma rapidly acquires a floating potential due to the imbalance between electron and ion collection. In the orbital-motion-limited (OML) regime~\cite{OML1926}, where the Debye length exceeds the droplet radius, collection is governed by collisionless single-particle orbits. The resulting charge is $q_d = 4\pi\epsilon_0 r_d \varphi$, where $\varphi$ denotes the surface potential.

For a collisionless Maxwellian plasma and in the absence of magnetic-field effects, the collected currents are given by
\begin{eqnarray}
\label{eq:Ie}
I_e &=& -\,\pi r_d^2\,n_e\,\sqrt{\frac{8k_B T_e}{\pi m_e}}\,
\exp\!\left(\frac{e\varphi}{k_B T_e}\right), \\[4pt]
\label{eq:Ii}
I_i &=& \phantom{-}\pi r_d^2\,n_i\,\sqrt{\frac{8k_B T_i}{\pi m_i}}\,
\left(1 - \frac{e\varphi}{k_B T_i}\right),
\end{eqnarray}
Under typical edge plasma conditions, the floating potential is negative, which suppresses electron collection and enhances ion collection.

For hot liquid metals, thermionic emission contributes an additional outward electron flux. This process is modeled using Richardson–Dushman emission with a size-corrected work function,
\begin{eqnarray}
\label{eq:Jth}
J_\mathrm{th} &=& A\,T_d^{\,2}\,\exp\!\left(-\frac{W(r_d)}{k_B T_d}\right),\\
W(r_d) &=& W^\infty - \frac{e^2}{16\pi\epsilon_0 r_d},
\end{eqnarray}
In these expressions, $A$ denotes the Richardson constant, $T_d$ is the droplet temperature, and the second term accounts for the Schottky-like reduction of the effective barrier for a finite-radius emitter. The steady-state surface potential is determined from the current balance
\begin{equation}
\label{eq:current_balance}
I_e + I_i + 4\pi r_d^2\,J_\mathrm{th} = 0,
\end{equation}
which is solved numerically to obtain $\varphi$. The droplet charge is subsequently given by
\begin{equation}
\label{eq:qd}
q_d = 4\pi\epsilon_0 r_d\,\varphi,
\end{equation}
This approach is consistent with the treatment of the droplet as a conducting sphere~\cite{2001JVSTA..19.2533S}.

This model employs a quasi-steady approximation: at each time step, the local plasma moments $(n_e,n_i,T_e,T_i)$ and droplet state $(r_d,T_d)$ define an instantaneous floating potential. The charge is recalculated at every step to track the evolving radius. This approximation is valid when the charging time is short relative to the relevant dynamical time scales and the sheath-edge distribution remains near-Maxwellian. Extensions to drifting plasmas and multi-charge-state ions follow standard treatments~\cite{DUSTT}.

  \subsection{Droplet--surface collisions}
  \label{sec:surfaceCollide}                                     
A negatively charged droplet approaching the wall must overcome the sheath electrostatic barrier. Contact requires a normal kinetic energy exceeding the sheath potential energy, corresponding to a threshold speed ${\sim}\sqrt{Z_d T_{e,\mathrm{sheath}}/ m_d}$. Slower droplets are reflected by the sheath field.

Upon wall contact, several collision models are available: (i) \textit{specular} reflection, which reverses the normal velocity component; (ii) \textit{diffuse} reflection, which re-emits the droplet from a cosine-law Maxwellian at wall temperature $T_w$ with sticking probability $\alpha\in[0,1]$; (iii) an energy- and angle-dependent accommodation model $\alpha(\theta,E)$ that interpolates between specular and diffuse limits using tabulated data; and (iv) a PMI model with pre-tabulated BCA reflection and sputtering yields $R(\theta,E)$ and $Y(\theta,E)$. The purely absorbing limit ($\alpha=1$) is the \textit{vanish} model. In the results presented here, all droplets that reach a surface are absorbed (vanish), while SOLPS-ITER via EIRENE handles sputtering and reflection for background plasma species.

\subsection{Evaporation model}
\label{sec:evaporationModel}

Under divertor conditions, plasma serves as the primary heat source, surpassing gas-phase convection. Therefore, an energy-balance evaporation model is employed, as described in~\cite{Crowe2011Multiphase,DUSTT}.

For a spherical droplet with uniform temperature and low Reynolds number ($\mathrm{Re}_d \ll 1$), the coupled mass and energy balances are given by
\begin{eqnarray}
\label{eq:radius_evol}
4\pi r_d^2 \rho_L \,\frac{dr_d}{dt} &=& -\,\dot{m}_d, \\[4pt]
\label{eq:temp_evol}
\frac{4}{3}\pi r_d^3 \rho_L C_p \,\frac{dT_d}{dt}
&=& Q_g - \dot{m}_d L_\mathrm{vap},
\end{eqnarray}
where $\rho_L$ is the liquid density, $C_p$ the specific heat capacity,
$T_d$ the droplet temperature, and $L_\mathrm{vap}$ the latent heat of vaporization.

The incident heat flux $Q_g$ represents the total energy transfer from the plasma and is decomposed as
\begin{equation}
Q_g = Q_e + Q_i + Q_{\rm rad} + Q_{\rm n},
\end{equation}
Here, $Q_e$ and $Q_i$ denote the electron and ion heat fluxes to the droplet surface, $Q_{\rm rad}$ represents the radiative heat load, and $Q_{\rm n}$ accounts for neutral collisions. In this study, $Q_g$ is obtained directly from the total heat-flux density calculated by SOLPS-ITER, which incorporates all four contributions. In edge or scrape-off layer (SOL) plasmas, electron heat flux generally dominates due to high electron thermal velocity and sheath acceleration, whereas ion contributions become significant at elevated flow rates or in detached regimes.

The mass loss rate $\dot{m}_d$ is evaluated using an energy-limited evaporation closure appropriate for plasma-exposed droplets,
\begin{equation}
\dot{m}_d = \frac{Q_g}{L_\mathrm{vap} + C_p (T_b - T_d)},
\end{equation}
Here, $T_b$ denotes the boiling temperature of lithium. When $T_d < T_b$, the absorbed energy increases the droplet temperature. Once $T_d$ approaches $T_b$, the majority of the incoming heat is utilized for phase change, consistent with the classical $d^2$-law. As the droplet radius decreases, both the collection area and capacitance are altered, which in turn modifies the heat flux and surface potential. Concurrently, evaporation introduces neutral lithium into the plasma, affecting local density and radiation. The feedback loop between OpenEdge and SOLPS-ITER that accounts for these interactions is detailed in Section~\ref{sec:two-way-coupling}.

In the isotropic limit, the net recoil resulting from evaporation is zero. However, in SOL and divertor plasmas, the heat flux is primarily aligned with the magnetic field, leading to preferential evaporation on the plasma-facing hemisphere. The resulting recoil, often referred to as the ``rocket'' force, is given by
\begin{equation}
\mathbf{F}_{\mathrm{rocket}} = -\eta\,\dot{m}_d\,v_{\mathrm{evap}}\,\hat{\mathbf{n}}_q,
\label{eq:rocket}
\end{equation}
Here, $v_{\mathrm{evap}} = \sqrt{2k_BT_b/m_{\mathrm{Li}}}$ represents the thermal speed of evaporated atoms, $\hat{\mathbf{n}}_q$ denotes the heat-flux direction (approximately the local magnetic field direction), and $\eta\in[0,1]$ is an asymmetry parameter. The limiting cases are $\eta=0$ for isotropic evaporation with no recoil, and $\eta=1$ for fully one-sided evaporation. For millimeter-scale lithium droplets, internal thermal equilibration occurs rapidly compared to evaporation, indicating that intermediate values are appropriate. In this work, $\eta$ is treated as a free parameter, and values of $\eta=0$, $0.5$, and $1.0$ are examined.

Within the Strang-split scheme (see Section~\ref{sec:Implementation/Tests}), the rocket force is implemented as a velocity increment $\Delta\mathbf{v} = -\eta\,(\Delta m/m_d)\,v_{\mathrm{evap}}\,\hat{\mathbf{n}}_q$ during each evaporation half-step. For a $2.5\,\mathrm{mm}$ droplet with $v_{\mathrm{evap}}\approx 1600\,\mathrm{m\,s^{-1}}$, the cumulative velocity change can approach the initial droplet speed if a substantial fraction of the mass is lost.

The model assumes (i) spatially uniform droplet temperature, (ii) negligible internal convection, and (iii) continuum thermodynamic properties for liquid Li. These hold for millimeter-scale droplets in low-Knudsen-number edge plasmas, where the thermal diffusion time is short compared to the evaporation time.

\section{Implementation and Verification}
\label{sec:Implementation/Tests}

Operator splitting is applied to the droplet update over a single global timestep $\Delta t$. Evaporation, Lorentz forcing (including position advection), and non-Lorentz forcing are integrated in a symmetric sequence to preserve second-order accuracy in $\Delta t$ for the split map~\cite{channell1996introduction, Tuckerman, StreitzPRL2006, LAMMPS}. The velocity-space non-Lorentz terms, specifically gravity and drag, are solved using an implicit half-step method, while the Lorentz term and position advection are advanced concurrently using the Boris pusher.

Let $f(\boldsymbol{\xi},t)$ denote the droplet phase-space density. We write
\begin{equation}
\partial_t f + \mathcal{L} f = 0,
\end{equation}
with
\begin{equation}
\mathcal{L} = \mathcal{L}_{\rm evap}+\mathcal{L}_{\rm EM}+\mathcal{L}_{\rm slow},
\end{equation}
where
\begin{align}
\mathcal{L}_{\rm evap} f &= \dot m_{\rm evap}\,\partial_{m_d} f, \\
\mathcal{L}_{\rm EM} f &= \mathbf{v}\cdot\nabla_{\mathbf{x}} f + \frac{q}{m_d}(\mathbf{E}+\mathbf{v}\times\mathbf{B})\cdot\nabla_{\mathbf{v}} f, \\
\mathcal{L}_{\rm slow} f &= \bigl(\mathbf{g}+\mathbf{a}_{\rm th}+\mathbf{a}_{\rm drag}(\mathbf{v}-\mathbf{u})\bigr)\cdot\nabla_{\mathbf{v}} f.
\end{align}

The global split map is
\begin{align}
\label{eq:global-split-map}
\exp(\Delta t\mathcal{L}) &\approx
\exp\!\left(\tfrac{\Delta t}{2}\mathcal{L}_{\rm evap}\right)
\exp\!\left(\tfrac{\Delta t}{2}\mathcal{L}_{\rm slow}\right)
\nonumber\\
&\quad
\exp\!\left(\Delta t\mathcal{L}_{\rm EM}\right)
\nonumber\\
&\quad
\exp\!\left(\tfrac{\Delta t}{2}\mathcal{L}_{\rm slow}\right)
\exp\!\left(\tfrac{\Delta t}{2}\mathcal{L}_{\rm evap}\right).
\end{align}

For the non-Lorentz half-step, we solve
\begin{equation}
\frac{d\mathbf{v}}{dt}
=
\mathbf{g}+\mathbf{a}_{\rm th}
+\mathbf{a}_{\rm drag}\!\left(\mathbf{v}_{\rm rel}\right),
\qquad
\mathbf{v}_{\rm rel}
=
\mathbf{v}-\mathbf{u},
\end{equation}
Coefficients are held constant over the substep. The half-step update from $\mathbf{v}^n$ to $\mathbf{v}^{n+1/2}$ is given by
\begin{equation}
\label{eq:halfkick}
\mathbf{v}^{n+1/2}
=
\mathbf{v}^{n}
+\frac{\Delta t}{2}
\left[
\mathbf{a}^{n}
+\mathbf{a}_{\rm drag}\!\left(\mathbf{v}^{n+1/2}-\mathbf{u}^{n}\right)
\right],
\end{equation}
Here, $\mathbf{a}^n = \mathbf{g}+\mathbf{a}_{\rm th}^n$ and $\mathbf{a}_{\rm drag}=\mathbf{F}_{\rm drag}/m_d$. In the case of linear drag, $\mathbf{a}_{\rm drag}=-\nu_E\mathbf{v}_{\rm rel}$, so Eq.~\eqref{eq:halfkick} admits a closed-form exponential solution. For nonlinear drag, the equation is solved using Newton iteration.

The Lorentz update in Eq.~\eqref{eq:global-split-map} is performed using the Boris method, which consists of an electric half-kick, magnetic rotation, and a second electric half-kick. If evaporation induces a significant change in $r_d$ within a half-step, the non-Lorentz kicks are subcycled:
\begin{equation}
\Delta t_{\rm sub} \le \min\!\left(\eta_\nu\,\nu_E^{-1},\,\eta_r\,\frac{r_d}{|\dot r_d|}\right).
\end{equation}

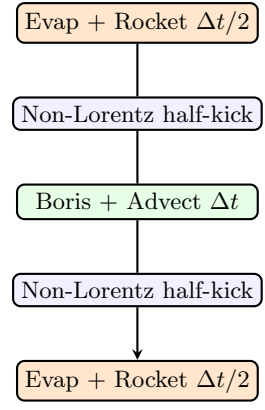
\begin{figure}[!t]
  \centering
  \begin{tikzpicture}[
      >=stealth,
      node distance=7mm,
      box/.style={draw, rounded corners=1mm, minimum width=32mm,
                  inner sep=3pt, font=\small},
      evap/.style={box, fill=orange!20},
      kick/.style={box, fill=blue!6},
      boris/.style={box, fill=green!10},
      every path/.style={thick}
    ]
    \node[evap]  (e1) {Evap + Rocket $\Delta t/2$};
    \node[kick,  below=of e1] (k1) {Non-Lorentz half-kick};
    \node[boris, below=of k1] (bo) {Boris + Advect $\Delta t$};
    \node[kick,  below=of bo] (k2) {Non-Lorentz half-kick};
    \node[evap,  below=of k2] (e2) {Evap + Rocket $\Delta t/2$};

    \draw[->] (e1) -- (k1) -- (bo) -- (k2) -- (e2);
  \end{tikzpicture}
  \caption{Time-integration scheme for a droplet subject to drag, gravity, evaporation, rocket recoil, and Lorentz forces. Evaporation (with rocket velocity kick) and non-Lorentz forces (drag, gravity) are applied as symmetric half-steps surrounding the Boris pusher, which advances both Lorentz rotation and position advection over the full timestep.}
  \label{fig:time-splitting-scheme}
\end{figure}

\subsection{Verification of the evaporation model}

The evaporation implementation is first verified in isolation ($\mathbf{g}=\mathbf{0}$, $\nu_E=0$, $\mathbf{u}=\mathbf{0}$), advancing only $\{r_d(t), m_d(t), T_d(t)\}$ under a uniform heat flux of $Q_s=50~\mathrm{MW\,m^{-2}}$. Three initial radii are examined: $r_{d0}=(1.5, 2.5, 3.5)~\mathrm{mm}$, with $T_{d0}=773.15~\mathrm{K}$.

\begin{figure}[!t]
\centering
\includegraphics[width=1\linewidth]{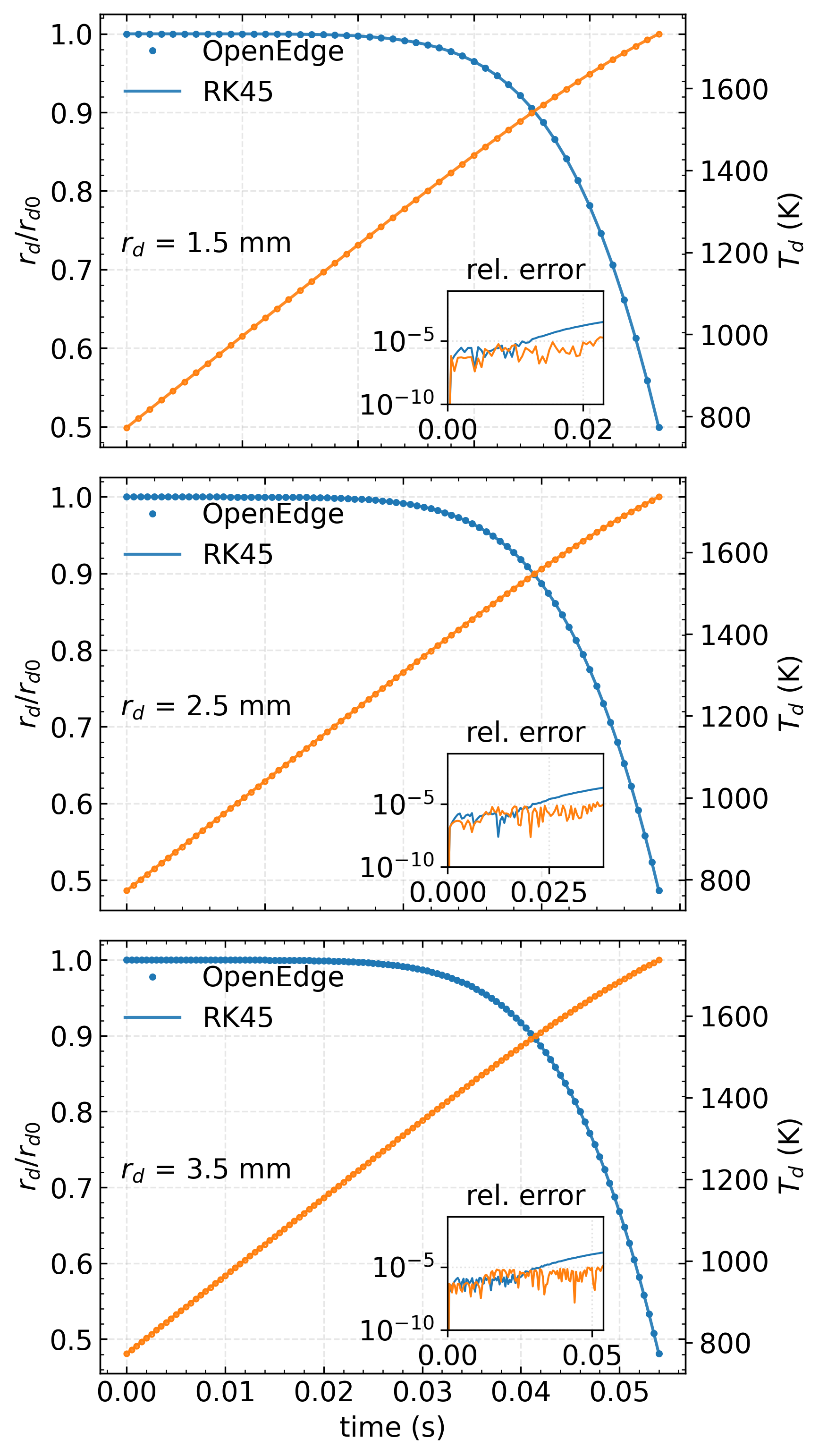}
\caption{Evaporation-only verification against RK45 for three droplet sizes
($r_{d0}=1.5,\ 2.5,\ 3.5~\mathrm{mm}$) at uniform heat flux
$Q_s=50~\mathrm{MW\,m^{-2}}$. Markers: OpenEdge; solid lines: RK45.
Left axis: $r_d/r_{d0}$, right axis: $T_d$.
Insets show relative error snippets for radius and temperature.}
\label{fig:evap-only}
\end{figure}

The results are compared against an independent RK45 integration of the same ordinary differential equation system, employing the same Antoine vapor-pressure closure and latent-heat coupling. As shown in Fig.~\ref{fig:evap-only}, the two solutions overlap throughout the simulation. Maximum differences remain below $5\times10^{-7}~\mathrm{m}$ for radius and $1.6\times10^{-2}~\mathrm{K}$ for temperature, with no systematic error growth observed near rapid shrinkage at late times.

\begin{figure*}[t]
\centering
\includegraphics[width=0.9\linewidth]{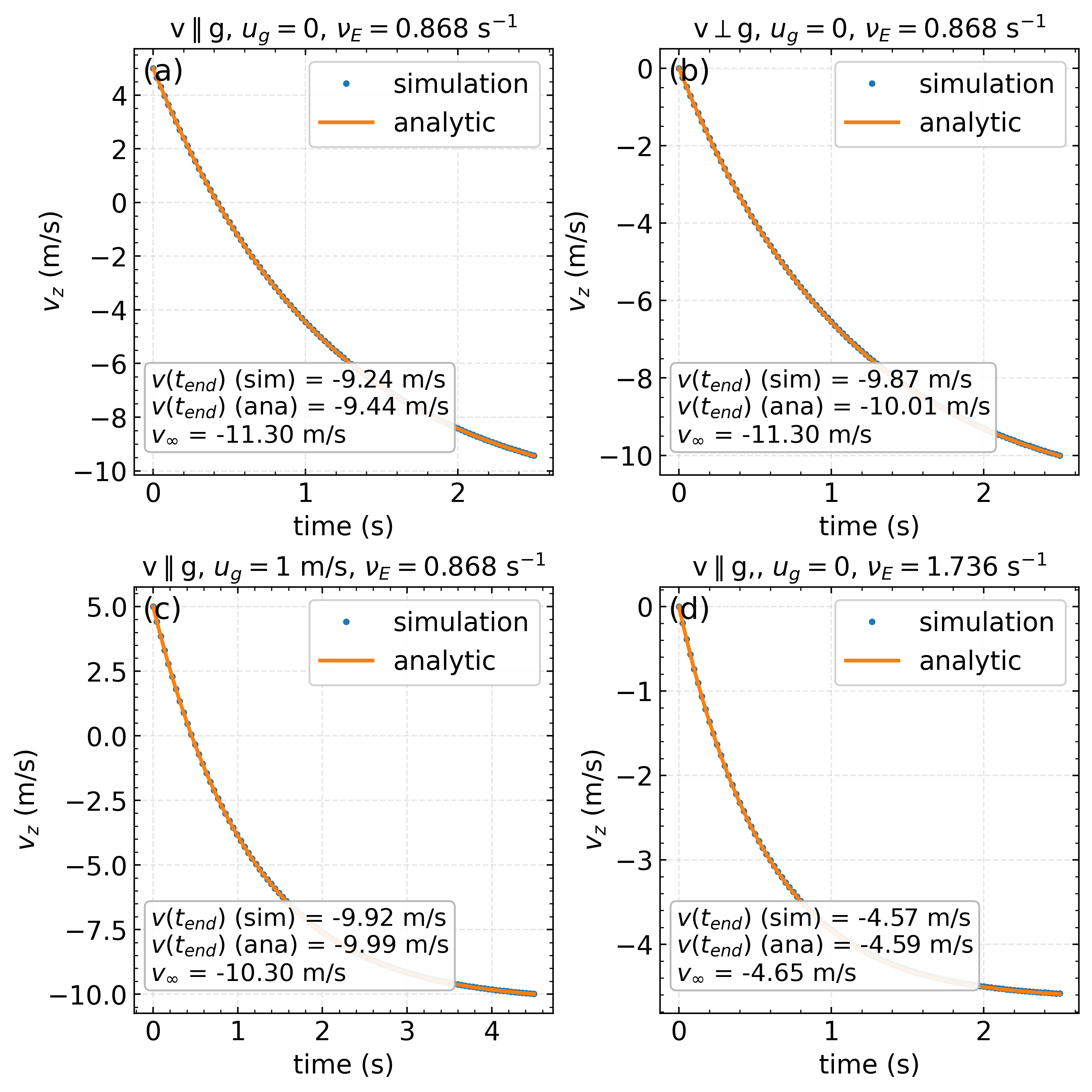}
\caption{Drag-gravity verification for a \(2.5~\mathrm{mm}\) lithium droplet.
Dotted lines: OpenEdge; solid lines: closed-form solution.
All panels use \(T_d=773.15~\mathrm{K}\), \(r_d=2.5~\mathrm{mm}\), \(m_d=3.495\times10^{-5}~\mathrm{kg}\), and no evaporation.
Panels show:
(a) \(v\!\parallel\!g\), \(u_g=0\), \(\nu_E=1.736\times10^{-2}~\mathrm{s}^{-1}\), \(v_{z,0}=+5~\mathrm{m\,s}^{-1}\);
(b) \(v\!\perp\!g\), \(u_g=0\), same \(\nu_E\);
(c) \(v\!\parallel\!g\), \(u_g=1~\mathrm{m\,s}^{-1}\), same \(\nu_E\);
(d) \(v\!\parallel\!g\), \(u_g=1~\mathrm{m\,s}^{-1}\), higher drag \(\nu_E=3.472\times10^{-2}~\mathrm{s}^{-1}\).
Insets list \(v(t_{\mathrm{end}})\), \(v_{\infty}\), and clarify that \(t_{\mathrm{end}}/\tau<1\) for these runs.}
\label{fig:gravity-drag-validation}
\end{figure*}

\begin{figure}[!t]
\minipage{0.25\textwidth}
  \includegraphics[width=1\linewidth]{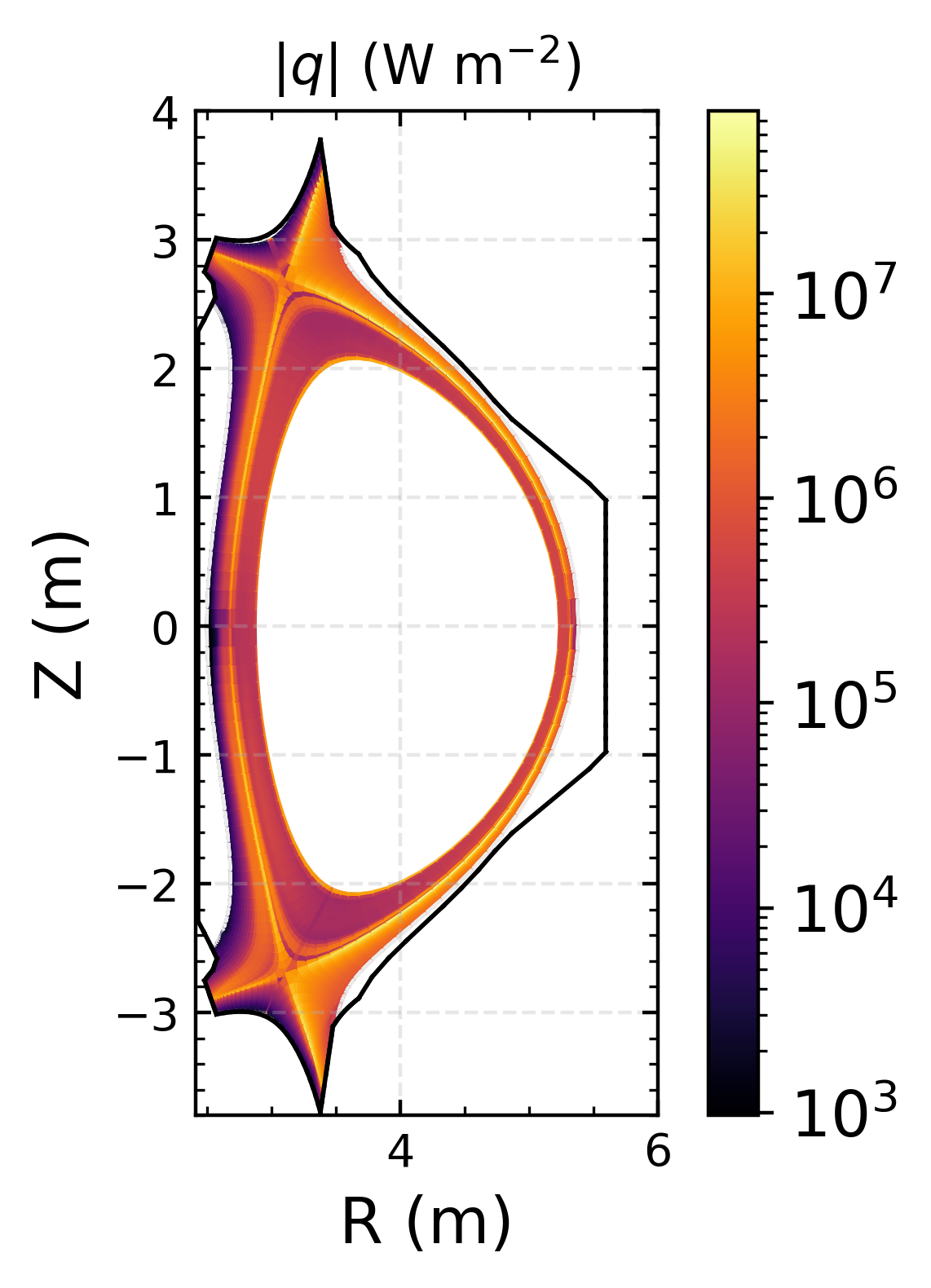}
\endminipage
\minipage{0.25\textwidth}
  \includegraphics[width=1\linewidth]{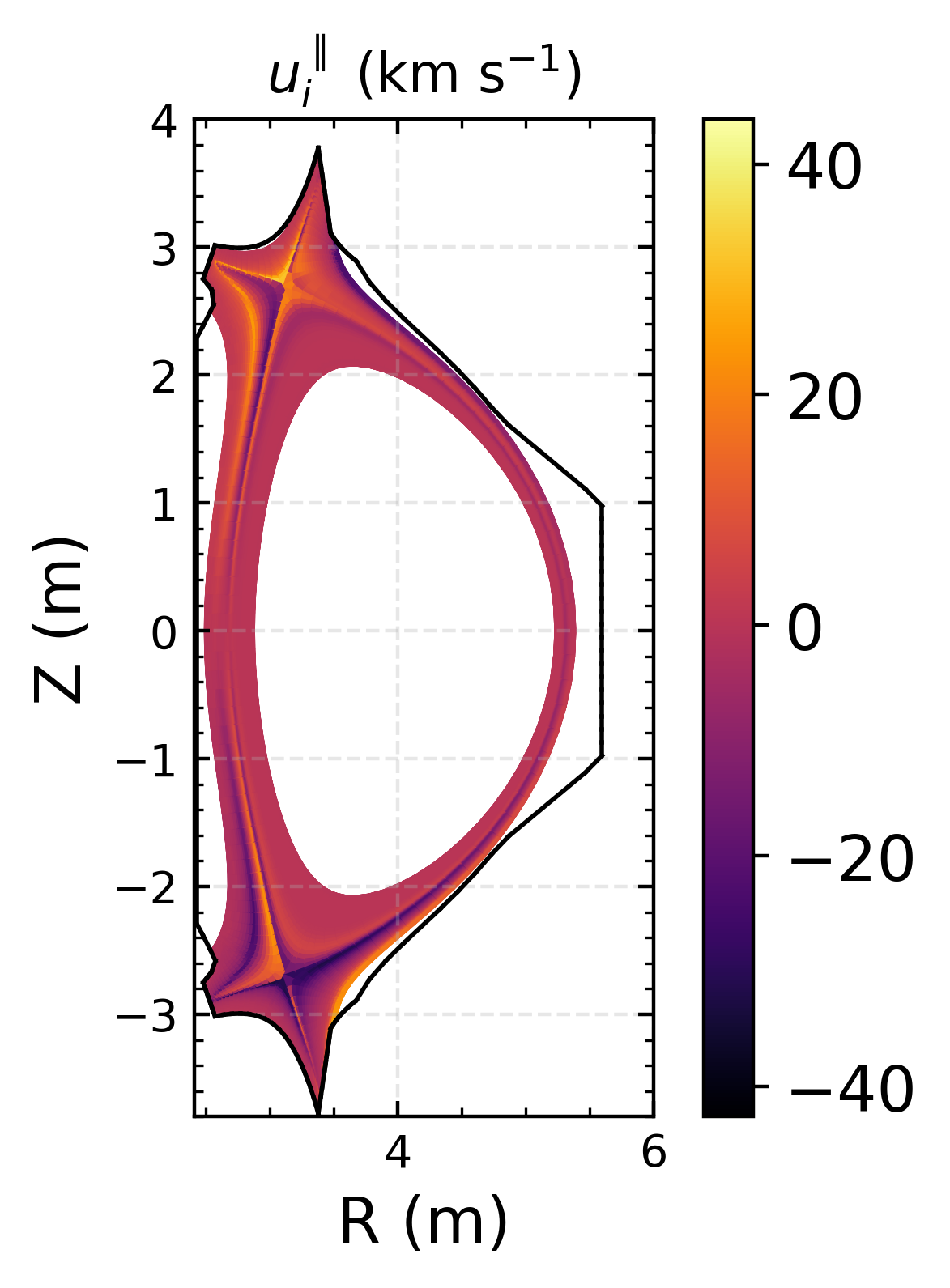}
\endminipage
  \caption{Total heat-flux density and main ion parallel flow velocity
    computed from SOLPS-ITER.}
  \label{fig:plasmaProfiles}
\end{figure}

\subsection{Verification of drag and gravity implementation}
\label{sec:grav-epstein-verification}

The gravity-drag update is benchmarked against the closed-form solution for a uniform plasma with constant $\nu_E$:
\begin{eqnarray}
v_z(t) &=& (v_0-v_\infty)e^{-\nu_E(t-t_0)}+v_\infty,
\label{eq:grav-epstein-update}
\end{eqnarray}
Here, $v_0$ denotes the initial velocity and $v_\infty=u_g+g/\nu_E$ represents the terminal value. A single droplet is used with $T_d=773.15~\mathrm{K}$, $r_d=2.5~\mathrm{mm}$, $m_d=3.495\times10^{-5}~\mathrm{kg}$, and $\Delta t=10^{-5}\,\mathrm{s}$, with evaporation disabled.

Figure~\ref{fig:gravity-drag-validation} presents a comparison between simulation results (dotted) and analytic solutions (solid) for four cases:
(a) $v\parallel g$, $u_g=0$, $\nu_E=1.736\times10^{-2}~\mathrm{s^{-1}}$, $v_z(0)=+5~\mathrm{m\,s^{-1}}$;
(b) $v\perp g$, $u_g=0$, same $\nu_E$;
(c) $v\parallel g$ with background drift $u_g=1~\mathrm{m\,s^{-1}}$, same $\nu_E$;
and (d) $v\parallel g$ with higher density, $\nu_E=3.472\times10^{-2}~\mathrm{s^{-1}}$.
For this droplet size, $\nu_E$ is small and the simulated intervals remain in the transient regime ($t_{\mathrm{end}}/\tau<1$, $\tau=1/\nu_E$), so $v_\infty$ is not reached. The normalized collapse $\tilde v=(v-v_\infty)/(v_0-v_\infty)$ versus $\tilde t=\nu_E(t-t_0)$ is also verified, confirming the accuracy of the drag-gravity update.

\subsection{Verification of the rocket force}
  \label{sec:rocket-verification}
\begin{figure*} 
\centering
\includegraphics[width=1\linewidth]{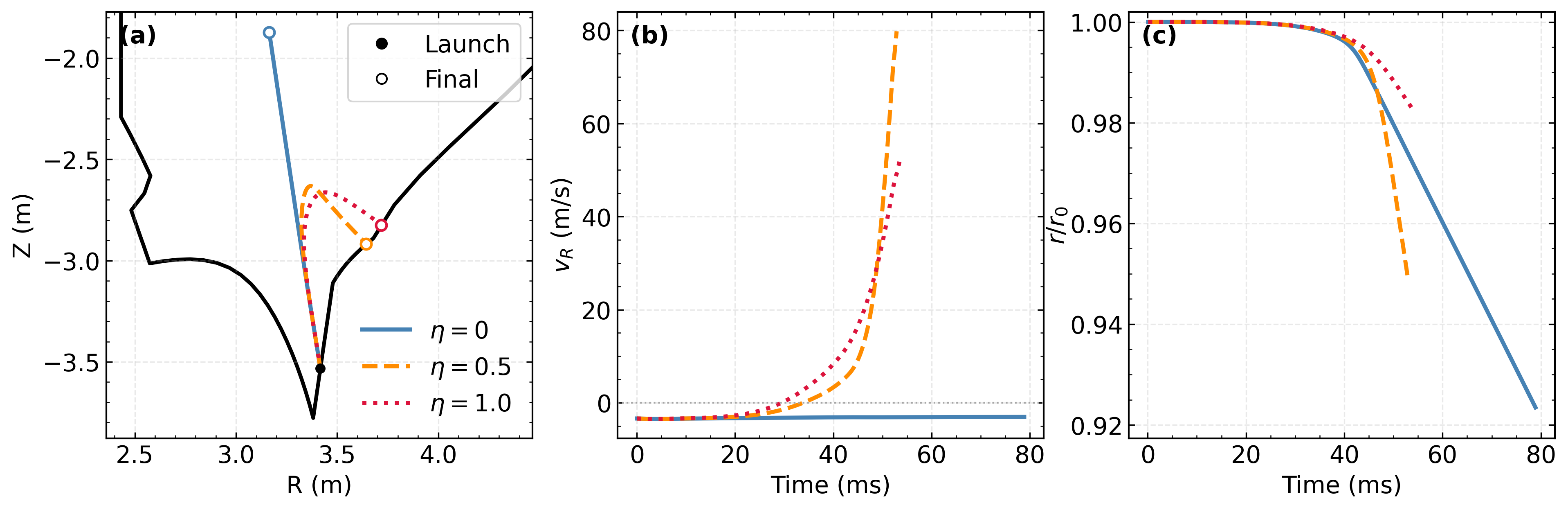}
\caption{Rocket-force verification for a $2.5\,\mathrm{mm}$ lithium droplet
launched from the outer divertor into a CAT SOLPS-ITER background plasma. Three asymmetry parameters are compared:
$\eta=0$ (no recoil), $0.5$, and $1.0$ (fully one-sided evaporation).
(a)~Trajectories in the $(R,Z)$ plane; filled circle marks the launch point, open circles mark the final position before wall absorption.
(b)~Radial velocity $v_R(t)$.
(c)~Normalized radius $r_d/r_{d0}$.}
\label{fig:rocket}
\end{figure*}

The rocket force is verified by launching a single $2.5\,\mathrm{mm}$ lithium droplet from the outer divertor into the CAT SOLPS-ITER background plasma, scanning $\eta=0$, $0.5$, and $1.0$. Figure~\ref{fig:rocket} displays (a) the trajectories in the $(R,Z)$ plane, (b) the radial velocity $v_R(t)$, and (c) the normalized radius $r/r_0$.

For $\eta=0$, the droplet follows a gravity-dominated arc with $v_R\approx  -3\,\mathrm{m\,s^{-1}}$, striking the core boundary wall after approximately $79\,\mathrm{ms}$. When the rocket force is activated ($\eta>0$), the radial drift reverses within about $10\,\mathrm{ms}$: $v_R$ reaches $+80\,\mathrm{m\,s^{-1}}$ for $\eta=0.5$ and $+49\,\mathrm{m\,s^{-1}}$ for $\eta=1.0$, ejecting the droplet outward into the wall at around $53\,\mathrm{ms}$. The lower terminal velocity for $\eta=1.0$ reflects the stronger recoil, which pushes the droplet out of the high-heat-flux zone sooner, thereby reducing subsequent evaporation and acceleration. This is confirmed by the radius histories: $\eta=0.5$ loses roughly 5\% of its initial radius before wall contact, compared to less than 2\% for $\eta=1.0$. The early-time minimum radial velocity $v_{R,\mathrm{min}}\approx-3.47\,\mathrm{m\,s^{-1}}$ is identical across all three cases, confirming that the rocket force contributes only after evaporation becomes significant.

\section{Parametric Studies and Discussions}
\label{sec:Results}
\subsection{Parametric Analysis}

\begin{figure*}
\centering
\includegraphics[width=1\linewidth]{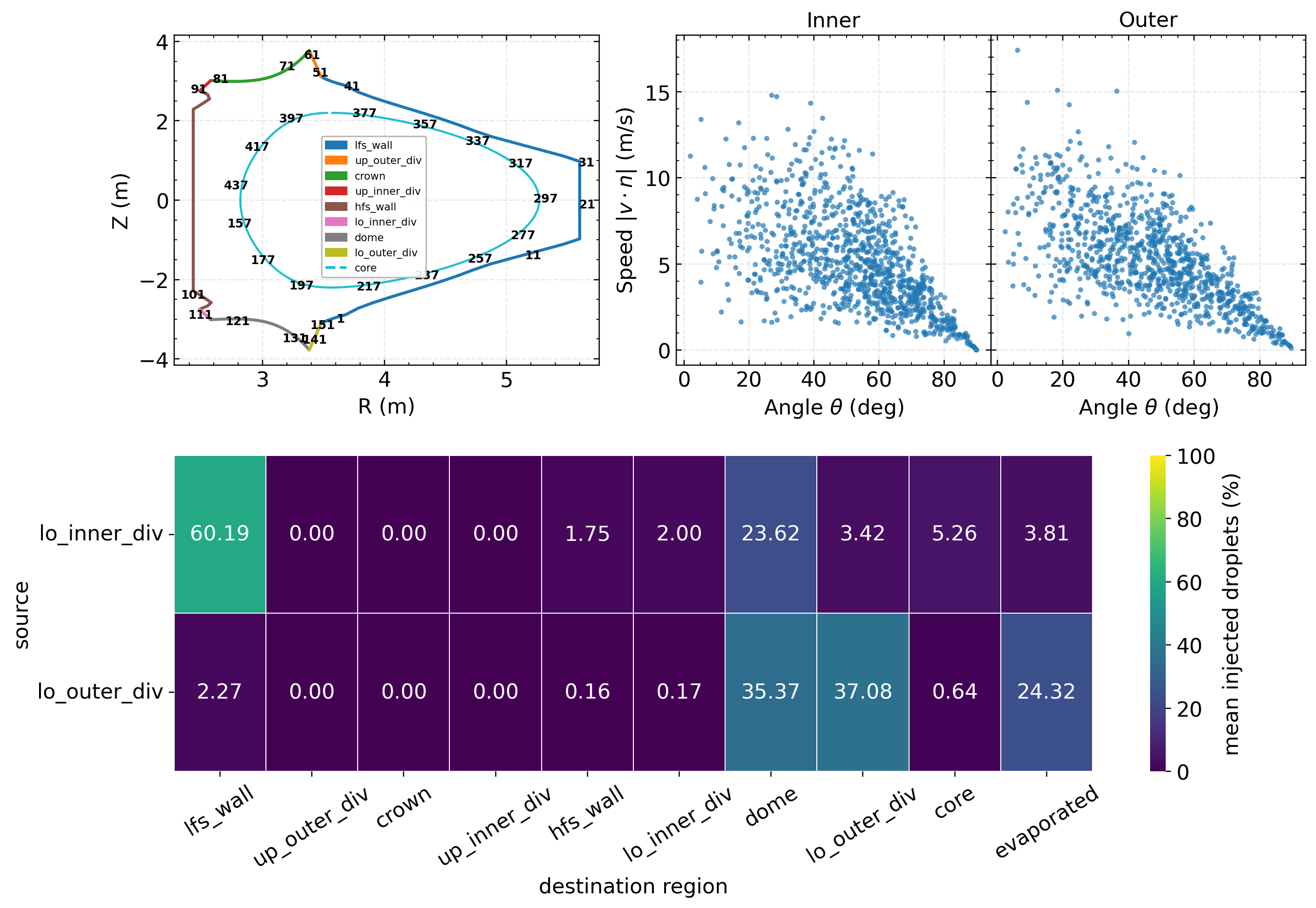}
\caption{
Region-resolved droplet transport statistics in a SOLPS-ITER plasma background.
Top-left: Tokamak poloidal cross-section in $(R,Z)$ showing the wall segmentation used for surface tallies (LFS wall, HFS wall, divertor plates, dome, crown, and core boundary).
Top-right: Distribution of injected droplet speed magnitude $|\mathbf{v}\cdot\mathbf{n}|$ and injection angle $\theta$ with respect to the local surface normal for inner and outer divertor sources.
Bottom: Row-normalized transfer matrix (in \%) representing the fate of droplets launched from each source region (rows) and their final destination region (columns), including redeposition on specific wall tiles, core penetration, and evaporation losses.
The matrix normalization ensures each row sums to 100\%, allowing direct comparison of redeposition pathways between inner and outer divertor injection regions under identical plasma conditions.
}
\label{fig:TransferMatrix}
\end{figure*}

Three representative droplet radii are considered,
$(1.5,\;2.5,\;3.5)~\mathrm{mm}$, with ensembles injected from both the inner and outer divertor. The droplet injection procedure is as follows:
\begin{enumerate}
\item The first-wall contour in the $(R,Z)$ poloidal plane is discretized into $156$ boundary segments, preserving the same segmentation used in SOLPS-ITER. Each segment, together with the core contour, is treated as an absorbing surface.
(Fig.~\ref{fig:TransferMatrix}, top left). The core boundary is similarly treated as an absorbing surface.
\item Droplets are emitted only from selected inner and outer divertor segments, with initial positions placed a few micrometers above the local surface.
\item For each injection event, the velocity is sampled using a Bird-type flux-weighted half-Maxwellian (normal component) with Maxwellian tangential components~\cite{Bird1994}. The thermal scale is $v_{\mathrm{scale}}=\sqrt{2k_B T/m_d}$, where $T$ is the local surface temperature.
\item The emission direction follows the Knudsen cosine law, $dN/d\Omega \propto \cos\theta$, with $\theta\in(0^\circ,90^\circ)$ measured from the surface normal. This is the standard DSMC emission model~\cite{Bird1994} and preferentially samples near-normal trajectories, consistent with surface-instability-driven ejection.
\end{enumerate}

\begin{figure}[!t]
\centering
\includegraphics[width=1\linewidth]{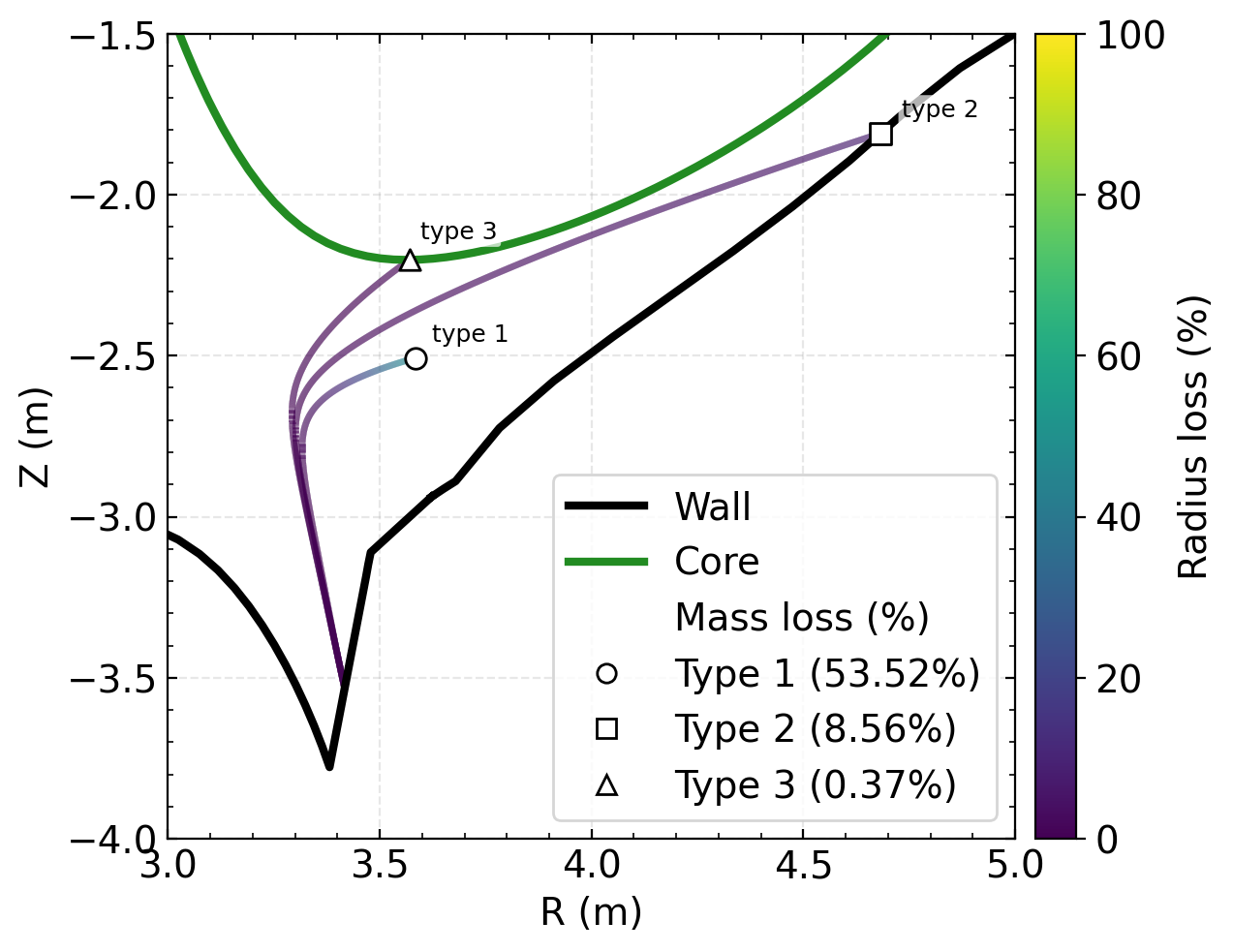}
\caption{Poloidal droplet trajectories colored by cumulative mass loss (
initial velocity $\mathbf{v}_0=(-3.40,\ 21.07,\ -2.52)\ \mathrm{m\,s^{-1}}$. The final mass-loss fraction decreases monotonically with increasing initial radius, indicating enhanced survival and deeper penetration for larger droplets.}
\label{fig:droplet_traj_massloss}
\end{figure}

We launch $10^5$ droplets from each divertor leg, distributed across all emitting segments. Droplets are tracked under gravity, collisional drag, electric charging, and evaporation until they either lose $90\%$ of their mass or impact a surface. The time step is $\Delta t=10^{-5}$~s. The plasma background and heat flux profiles are taken from SOLPS-ITER; Figure~\ref{fig:plasmaProfiles} shows the parallel flow and heat flux used.

Figure~\ref{fig:droplet_traj_massloss} presents three representative droplet trajectories in the poloidal $(R,Z)$ plane, where the trajectory color indicates cumulative mass loss (in percent). The three cases correspond to initial droplet radii $r_{d0}
=1.5,\ 2.5,\ 3.5~\mathrm{mm}$ (Types 1–3, respectively). All droplets are initialized with the same velocity vector,
\begin{eqnarray}
\mathbf{v}_0=(v_R,v_Z,v_\phi)=(-3.40,\ 21.07,\ -2.52)\ \mathrm{m\,s^{-1}}
\end{eqnarray}
corresponding to a total speed $|\mathbf{v}_0|\approx 21.5 \, \mathrm{m\,s^{-1}}$ and a poloidal launch angle of approximately $9.2^\circ$ from the local normal. A clear size-dependent trend is observed: the smallest droplet (Type 1) undergoes the
largest fractional mass loss, while the largest droplet (Type 3) retains most of its mass over a similar path length. This representation directly links orbit topology and evaporation, highlighting initial size as a primary control
parameter for lithium droplet survivability and penetration into the core.

Figure~\ref{fig:TransferMatrix} shows the droplet transfer matrices, which quantify the transport of droplets launched near the divertor across the surrounding plasma-facing components (PFCs). A clear hierarchy of transport patterns emerges across droplet sizes and injection locations.

\subsection{Coupling with SOLPS-ITER}
\label{sec:two-way-coupling}

\begin{figure*}
\centering
\includegraphics[width=1\linewidth]{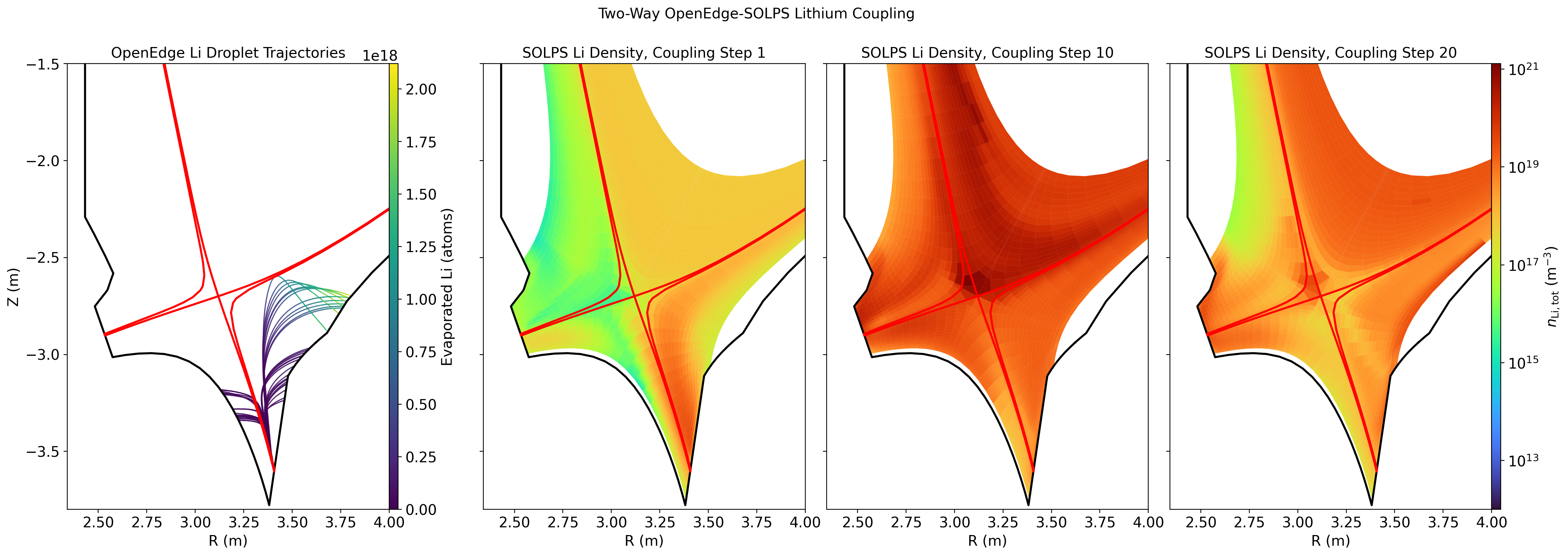}
\caption{%
Two-way OpenEdge-SOLPS coupling results for 50 lithium droplets with initial radius $r_0 = 2.5\,\mathrm{mm}$ injected from the outer divertor. Left: OpenEdge droplet trajectories colored by cumulative evaporated lithium atoms. Right: SOLPS-ITER total lithium density $n_{\mathrm{Li,tot}}$ at
coupling steps 1, 10, and 20, showing the redistribution of lithium as the plasma response evolves.%
  }
\label{fig:couplingResults}
\end{figure*}

OpenEdge employs a regular Cartesian mesh in $(R,Z)$, whereas SOLPS-ITER utilizes an irregular, flux-surface-aligned quadrilateral mesh. Communication between the two codes requires mapping the OpenEdge evaporation output onto the SOLPS grid. For each OpenEdge cell, the evaporated lithium is converted to a volumetric source,
\begin{equation}
S_{\mathrm{Li}}^{\mathrm{OE}}=\frac{\Delta N_{\mathrm{Li}}^{\mathrm{OE}}}{V_{\mathrm{OE}}\Delta t_{\mathrm{OE}}},
\end{equation}
Here, $\Delta N_{\mathrm{Li}}^{\mathrm{OE}}$ denotes the evaporated lithium per timestep and $V_{\mathrm{OE}}$ represents the cell volume. This field is mapped to SOLPS cell centers using nearest-cell interpolation to get the total source rate:
\begin{equation}
\dot N_{\mathrm{Li},ij}=S_{\mathrm{Li},ij}V_{ij},
\qquad (\mathrm{atoms\,s^{-1}}).
\end{equation}
The source history is coarse-grained into discrete time windows, with one \texttt{source2d} file generated per window and linked via chained \texttt{b2.sources.profile} files. No modifications to the SOLPS-ITER source code are required, as the driver communicates through the existing \texttt{source2d} interface and reads the plasma state from standard SOLPS-ITER output files.

In the simplest mode (one-way coupling), OpenEdge runs to completion in a fixed SOLPS background and its evaporation output is post-processed into time-dependent lithium source files.

To capture the plasma response self-consistently, this workflow is extended to an iterative two-way coupling implemented via a Python driver. In each coupling step $k$:
\begin{enumerate}
\item OpenEdge advances for $N_{\mathrm{OE}}$ timesteps in the current SOLPS background, accumulating the per-cell evaporated Li source $\Delta N_{\mathrm{Li}}^{(k)}$.
\item The source is mapped to the SOLPS mesh and converted to a particle source rate:
\begin{equation}
\dot{N}_{\mathrm{Li},ij}^{(k)} = \frac{\Delta N_{\mathrm{Li},ij}^{(k)}}{N_{\mathrm{OE}}\,\Delta t_{\mathrm{OE}}}.
\end{equation}
\item The source supplied to SOLPS is under-relaxed for stability:
\begin{equation}
\dot{N}_{\mathrm{Li},ij}^{\mathrm{used}} = \alpha\,\dot{N}_{\mathrm{Li},ij}^{(k)} + (1-\alpha)\,\dot{N}_{\mathrm{Li},ij}^{(k-1)},
\label{eq:underrelax}
\end{equation}
with relaxation parameter $\alpha$ chosen between $0$ and $1$. In this work, $\alpha=0.5$.
\item The relaxed source is written as a \texttt{source2d} file and SOLPS-ITER is advanced for $N_{\mathrm{SOLPS}}$ iterations.
\item The updated plasma state ($n_e$, $T_e$, $T_i$, $u_\parallel$) is interpolated onto a regular $(R,Z)$ grid and written as an \emph{HDF5} file for OpenEdge.
\item OpenEdge resumes from a restart file that preserves all particle states (positions, velocities, mass, radius, temperature), ensuring trajectory continuity across coupling steps.
\end{enumerate}

The current implementation exchanges only the particle source $\dot{N}_{\mathrm{Li}}$. The evaporation model also calculates the energy absorbed per cell, which can be mapped to SOLPS heat sinks $S_{he}$ and $S_{hi}$ using the same workflow. For isotropic evaporation ($\eta=0$), the net momentum deposited is zero. When the rocket force is active ($\eta>0$), anisotropic evaporation generates a momentum source that can also be exchanged.

\begin{figure}[!t]
\centering
\includegraphics[width=1\linewidth]{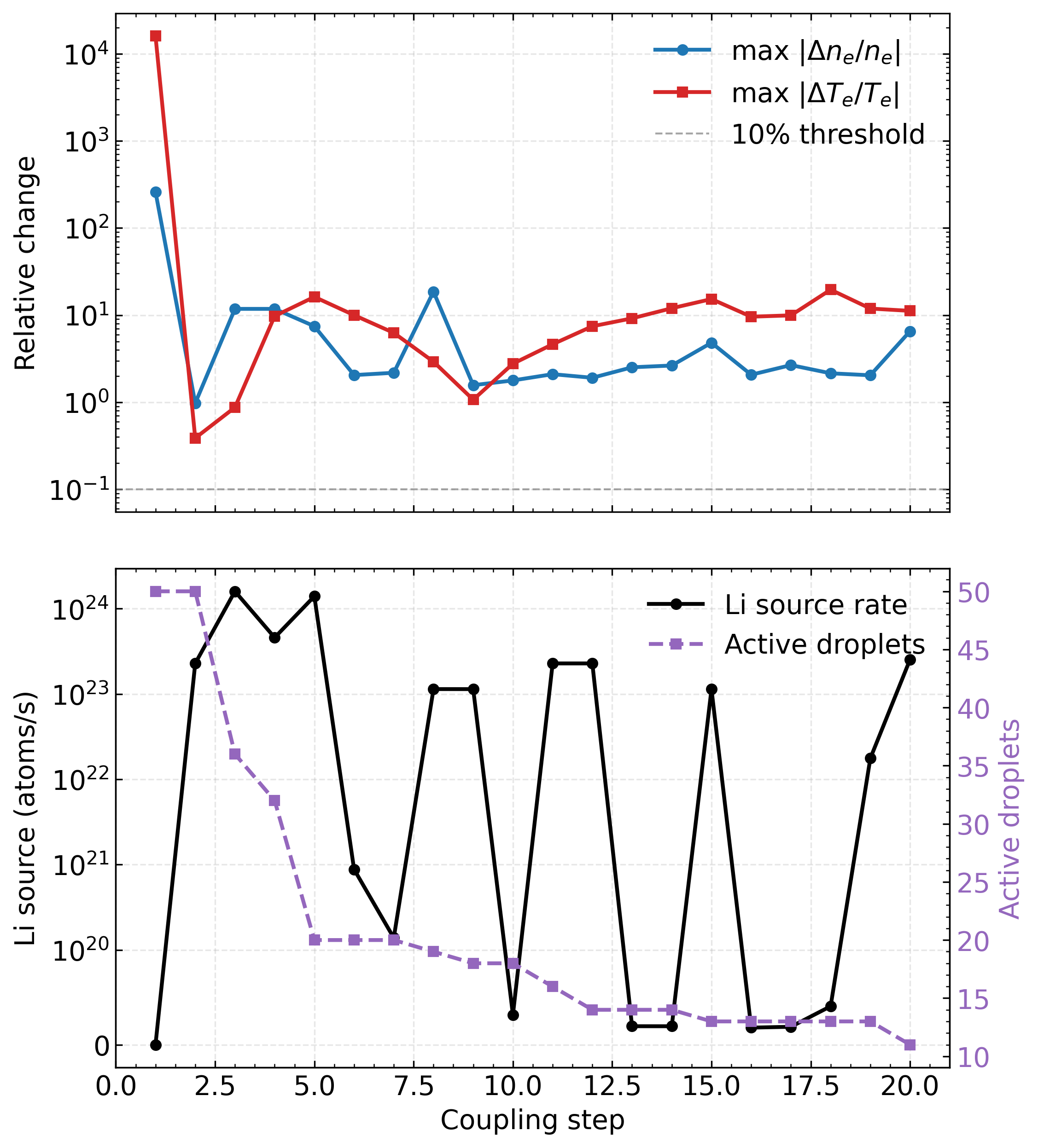}
\caption{Two-way OpenEdge--SOLPS coupling history for 50 lithium droplets with $r_0=2.5\,\mathrm{mm}$ and under-relaxation $\alpha=0.5$. Top: maximum relative changes in electron density and temperature between successive coupling steps, together with the 10\% reference level. Both quantities decrease sharply after the initial iteration but remain
above the 10\% level over the 20-step run. Bottom: lithium source rate (left axis) and number of active droplets (right axis) versus coupling step. The source remains strongly intermittent, while the active droplet population decreases monotonically as droplets impact surfaces or fully evaporate.}
\label{fig:coupling_convergence}
\end{figure}

As a demonstration, Fig.~\ref{fig:couplingResults} shows a two-way OpenEdge--SOLPS coupling run for 50 lithium droplets with initial radius $r_0=2.5\,\mathrm{mm}$ injected from the outer divertor. The left panel shows OpenEdge droplet trajectories colored by cumulative evaporated lithium, while the three right panels show the SOLPS total lithium
density $n_{\mathrm{Li,tot}}$ at coupling steps 1, 10, and 20. The plasma response changes rapidly during the first few coupling iterations and continues to evolve over the remainder of the run. Figure~\ref{fig:coupling_convergence} shows the corresponding coupling history for under-relaxation $\alpha=0.5$. The maximum relative changes $|\Delta n_e/
n_e|$ and $|\Delta T_e/T_e|$ decrease sharply after the initial iteration, but remain above the 10\% level throughout the 20-step run. The lithium source rate is highly intermittent, varying by several orders of magnitude as droplets traverse high heat-flux regions, while the number of active droplets decreases from 50 to 11 as droplets impact
plasma-facing surfaces or fully evaporate. In the final plotted state (coupling step 20), the total lithium density is concentrated near the inner divertor target close to the separatrix strike region, reaching values of order $10^{20}\,\mathrm{m}^{-3}$.
  
\section{Conclusions}

A Lagrangian lithium droplet model was implemented in the DSMC code OpenEdge, incorporating gravity, collisional ion drag, OML charging, energy-balance evaporation, and an anisotropic rocket recoil force within a Strang-split integrator. Verification against analytical drag-gravity solutions and independent RK45 evaporation integration demonstrates relative errors below $10^{-5}$.

The rocket force was verified by scanning $\eta=0$, $0.5$, and $1.0$ for a single droplet in a CAT SOLPS-ITER background. As $\eta$ increases, the radial drift transitions from inward, dominated by gravity, to outward, dominated by recoil. The resulting trajectory and mass-loss histories align with the expected physical behavior.

Simulations of ensembles of $10^5$ droplets launched from CAT divertor surfaces indicate a pronounced size dependence. The smallest droplets ($r_{d0}=1.5$~mm) lose approximately 53\% of their mass to evaporation, with about 5\% of inner-divertor droplets penetrating to the core. In contrast, the largest droplets ($r_{d0}=3.5$~mm) lose less than 1\% of their mass and redeposit on nearby tiles (Fig.~\ref{fig:TransferMatrix}). Only the smallest breakup products offer an efficient pathway for lithium transport to the core.

Both one-way and iterative two-way coupling between OpenEdge and SOLPS-ITER were demonstrated. The two-way framework exchanges lithium sources and updated plasma profiles at prescribed intervals, employing under-relaxation to ensure stability. In a 50-droplet demonstration case, density and temperature residuals converge to within 10--20\% after approximately 10 coupling iterations, with the lithium density reaching $n_{\mathrm{Li,tot}}\approx 1.8\times10^{18}$~m$^{-3}$ at the inner divertor strike point. Droplet states are maintained across coupling steps using restart files, and no modifications to the SOLPS-ITER source code are required.

Planned extensions include incorporating the evaporation energy sink in the SOLPS coupling, implementing tighter in-memory coupling to eliminate file input and output, and integrating a liquid-metal magnetohydrodynamic (MHD) surface code for self-consistent droplet generation from free-surface instabilities.

\section*{Acknowledgments}

\noindent This manuscript has been authored by UT-Battelle, LLC, with the US Department of Energy (DOE). This material is based upon work supported by the U.S. Department of Energy, Office of Science, Office of Fusion Energy Sciences under Award DE-AC05-00OR22725 as part of the SciDAC program. This research used resources of the Oak Ridge National Laboratory Research Cloud. The results are obtained with the help of the EIRENE package (see www.eirene.de) including the related code, data and tools \cite{Reiter2005TheCodes}.

\bibliography{paper}

\section*{Code and Data availability}

The particle simulation code (OpenEdge) is available in the repository:
\url{https://github.com/ORNL-Fusion/OpenEdge}.

SOLPS-ITER is available from the official ITER Organization repository:
\url{https://github.com/iterorganization/SOLPS-ITER}.

The scripts used for post-processing and droplet test cases are provided at:
\url{https://github.com/ORNL-Fusion/OpenEdge/tree/main/examples/test_droplet}.


\end{document}